\documentclass[letterpaper, twocolumn, prl]{revtex4-2}
\usepackage{graphicx, amsmath, amssymb, bm, titlesec, mathtools, braket, natbib}

\begin{document}

\title{Measurement of two-photon position-momentum
EPR correlations through single-photon
intensity measurements}

\author{Abhinandan Bhattacharjee, Nilakantha Meher and Anand K. Jha}

\email{akjha9@gmail.com}

\affiliation{Department of Physics, Indian Institute of
Technology Kanpur, Kanpur UP 208016, India}

\date{\today}

\begin{abstract}

The measurement of the position-momentum EPR correlations of a two-photon state is important for many quantum information applications ranging from quantum key distribution to coincidence imaging. However, all the existing techniques for measuring the position-momentum EPR correlations involve coincidence detection and thus suffer from issues that result in less accurate measurements. In this letter, we propose and demonstrate an experimental scheme that does not require coincidence detection for measuring the EPR correlations. Our technique works for two-photon states that are pure, irrespective of whether the state is separable or entangled. We theoretically show that if the pure two-photon state satisfies a certain set of conditions then the position-momentum EPR correlations can be obtained by doing the intensity measurements on only one of the photons. We experimentally demonstrate this technique for pure two-photon states produced by type-I spontaneous parametric down-conversion, and to the best of our knowledge, we report the most accurate measurement of position-momentum EPR correlations so far. 
\end{abstract}

\maketitle

If two photons are entangled in position and momentum variables, the product of the conditional position and momentum uncertainties of the individual photons becomes less than $0.5\hbar$, the minimum value allowed by the Heisenberg uncertainty relation. This violation of the conditional Heisenberg uncertainty relation is the signature of position-momentum EPR correlations \cite{einstein1935pr} and is a witness of the position-momentum entanglement. For two-dimensional variables, such as polarization, entanglement can be verified through the violations of  Bell's inequalities \cite{bell1964physics} and can be quantified through measures such as concurrence \cite{wootters1998prl}. However, for continuous variables, such as position-momentum, there is no prescription for quantifying entanglement. One can at best only verify entanglement, and the EPR-correlations measurements are the primary tool for that. In the past, several studies have used EPR-correlations measurements in the position-momentum variables in order to demonstrate position-momentum entanglement  \cite{d2004prl, howell2004prl,zhang2019optexp, o2005prl,leach2012pra, edgar2012natcom,reichert2018scirpt, moreau2014prl}. EPR-correlation measurements have also been used as witnesses of entanglement in many other continuous variables including time-energy \cite{khan2006pra,maclean2018prl,mei2020prl}, angle-orbital angular momentum (OAM) \cite{leach2010science}, radial position-radial momentum \cite{chen2019prl}, and quadrature phase-amplitude \cite{ou1992prl}.  More recently, even in entangled systems not consisting of photons, EPR-correlations measurements have become important tools for witnessing continuous-variable entanglement. These include macroscopic objects \cite{ockeloen2018nature}, Bose Einstein condensate \cite{fadel2018science}, and cold atoms \cite{josse2004prl}.

The demonstration of position-momentum EPR-correlations is very important for many applications such as quantum key distribution  \cite{almeida2005experimental}, quantum information processing \cite{dixon2012prl}, quantum metrology \cite{brida2010natphot}, coincidence imaging \cite{bennink2004prl,aspden2013njp}, and coincidence holography \cite{defienne2019entanglement}, since the efficiency of these applications relies on how accurately the EPR correlations could be measured. Therefore, it is very important to have a more accurate technique for measuring EPR-correlations. In the past few years, many schemes with increased accuracy have been demonstrated \cite{d2004prl, howell2004prl, zhang2019optexp, o2005prl,leach2012pra, edgar2012natcom,reichert2018scirpt, moreau2014prl, khan2006pra,maclean2018prl,mei2020prl, leach2010science, chen2019prl, ou1992prl}. However, all these methods involve coincidence detection,  implemented either by using two scanning single-photon detectors \cite{d2004prl}, or two scanning slits \cite{howell2004prl,zhang2019optexp}, or array of single-photon detectors \cite{o2005prl,leach2012pra}, or EMCCD cameras \cite{edgar2012natcom,reichert2018scirpt, moreau2014prl}. As a result, these measurement methods suffer from either too much loss of light, or strict alignment requirements, or multiple measurements, which adversely affect the accuracy of measurements. 

On the other hand, in the context of two-dimensional two-particle state, that is, the two-qubit states, it is known that if the state is pure, the entanglement quantifiers \cite{wootters1998prl, hill1997prl} can be measured by doing measurements on only one of the qubits \cite{walborn2006nature, cheng2016scirpt}, without requiring coincidence detection. Furthermore, even in the context of continuous variables, it is known that when the two-photon state is pure, several two-photon properties such as two-photon angular Schmidt spectrum \cite{Jha2011PRA,kulkarni2017natcom,kulkarni2018pra}, two-photon spatial Schmidt number \cite{pires2009pra}, and momentum correlations  \cite{hochrainer2017pnas} can be measured by doing intensity measurements on only one of the subsystems. These measurement schemes based on intensity detection provide much better accuracy than those based on coincidence detection. In this letter, utilizing the same physics, we propose a technique for measuring the position-momentum EPR-correlations that does not require coincidence detection. Our technique is based on measuring the intensities of only one of the subsystems, and it works for two-photon states that are pure, irrespective of whether the state is separable or entangled. We show that if a pure two-photon state satisfies a certain set of conditions then the position-momentum EPR correlations can be obtained by doing the intensity measurements on only one of the photons. We experimentally demonstrate our technique with pure two-photon states produced by type-I spontaneous parametric down-conversion (SPDC), and we obtain, to the best of our knowledge, the most accurate measurement of position-momentum EPR-correlations reported so far.

A pure state of two photons $\ket\Psi$ in the transverse momentum basis can be written as 
\begin{align}
\ket\Psi=\iint d\bm{p}_1 d\bm{p}_2 \psi(\bm{p}_1,\bm{p}_2) \ket{\bm{p}_1,\bm{p}_2}.\label{eq1}
\end{align} 
Here, $\bm{p}_1 \equiv (p_{1x}, p_{1y})$ and $\bm{p}_2 \equiv (p_{2x}, p_{2y})$ are the transverse momenta of the first and the second photon, respectively, $\ket{\bm{p}_1,\bm{p}_2}$ is the two-photon state vector, and $\psi(\bm{p}_1,\bm{p}_2)$ represents the two-photon transverse-momentum wavefunction. We note that if the two-photon wavefunction satisfies 
$\psi(\bm{p}_1,\bm{p}_2)=\psi(\bm{p}_1)\psi(\bm{p}_2)$ then it is separable, otherwise it is non-separable, or in other words, entangled. When the second photon is detected with transverse momentum $\bm{p}_2=0$, then the conditional momentum probability distribution $P(\bm{p}_1|\bm{p}_2=0)$ of the first photon is given by 
\begin{align}
P(\bm{p}_1|\bm{p}_2=0)=|\psi(\bm{p}_1,\bm{p}_2=0)|^2.\label{eq2}
\end{align} 
Now, the momentum cross-spectral density function of the first photon can be calculated as $W(\bm{p}_1,\bm{p'}_1)=\bra\Psi E^{(+)}_s(\bm{p}_1)E^{(-)}_1(\bm{p'}_1)\ket\Psi$ \cite{dixon2010pra}, 
where $E^{(+)}_1(\bm{p}_1)$ and $E^{(-)}_1(\bm{p'}_1)$ are the negative- and positive-frequency parts of the electric field operators respectively. For $\bm{p'}_1=-\bm{p}_1$, we have
\begin{align}
W(\bm{p}_1,-\bm{p}_1)=\iint \psi^{*}(\bm{p}_1,\bm{p}_2)\psi(-\bm{p}_1,\bm{p}_2)d\bm{p}_2.\label{eq3}
\end{align}
Next, we find that if the two-photon wavefunction satisfies the following condition  
\begin{align}
\!\!\!\!\! \psi^{*}(\bm{p}_1,\bm{p}_2)\psi(-\bm{p}_1,\bm{p}_2)\propto |\psi(\bm{p}_1,\bm{p}_2=0)\psi(\bm{p}_1=0,\bm{p}_2)|^2, \label{eq5}
\end{align} 
then using Eqs.~(\ref{eq2}) and ~(\ref{eq3}), one can show that
\begin{align}
W(\bm{p}_1,-\bm{p}_1) \propto P(\bm{p}_1|\bm{p}_2=0).  \label{eq6}
\end{align}
We note that the condition in Eq.~(\ref{eq5}) can be satisfied by both separable and inseparable pure two-photon wavefunctions. Eq.~(\ref{eq6}) is the main theoretical result of this letter. It states that as long as a two-photon state is pure, whether separable or entangled, and satisfies the condition in Eq.~(\ref{eq5}), the momentum cross-spectral density function of the first photon remains proportional to its conditional momentum probability distribution function. This implies that the standard deviations of $W(\bm{p}_1,-\bm{p}_1)$ and $P(\bm{p}_1|\bm{p}_2=0)$ are equal and that by measuring the standard deviation of $W(\bm{p}_1, -\bm{p}_1)$, one can obtain the standard deviation of $P(\bm{p}_1|\bm{p}_2=0)$. We denote the standard deviation of the conditional $x$-momentum of the first photon by $\Delta({p_{1x}|p_{2x}}=0)$. 

Now, by writing the two-photon wavefunction of Eq.~(\ref{eq1}) in the position basis and proceeding in the similar manner, we can show that if the two-photon position wavefunction $\psi(\bm{\rho}_1,\bm{\rho}_2)$ satisfies the condition 
\begin{align}
\!\!\!\!\! \psi^{*}(\bm{\rho}_1,\bm{\rho}_2)\psi(-\bm{\rho}_1,\bm{\rho}_2)\propto |\psi(\bm{\rho}_1,\bm{\rho}_2=0)\psi(\bm{\rho}_1=0,\bm{\rho}_2)|^2,\label{eq7}
\end{align}
where $\bm{\rho}_1\equiv (x_1, y_1)$ and $\bm{\rho}_2 \equiv (x_2, y_2)$ are the transverse position vectors of the first and the second photon, then the position cross-spectral density function $W(\bm{\rho}_1,-\bm{\rho}_1)$ of the first photon is proportional to its conditional position probability distribution function $P(\bm{\rho}_1|\bm{\rho}_2=0)$, that is, 
\begin{align}
W(\bm{\rho}_1,-\bm{\rho}_1)\propto P(\bm{\rho}_1|\bm{\rho}_2=0).\label{eq7new}
\end{align}
Thus, by measuring the standard deviation of $W(\bm{\rho}_1, -\bm{\rho}_1)$, one can obtain the standard deviation of $P(\bm{\rho}_1|\bm{\rho}_2=0)$. We denote the standard deviation of the conditional $x$-position of the first photon by $\Delta ({x_1|x_2}=0)$. We note that although the above analysis has been presented with respect to making measurements on the first photon, we obtain the same result even when analysed with the second photon. Now, it is known that if the two-photon wavefunction is separable then the product $U$ of the conditional uncertainties satisfies the Heisenberg uncertainty relation, that is,
\begin{align}
U\equiv \Delta ({x_1|x_2}=0)\Delta({p_{1x}|p_{2x}}=0)	\geqslant 0.5\hbar. \label{Heisenberg-4}
\end{align}
However, a violation of this inequality implies that the two-photon wavefunction is non-separable and that the two photons are entangled having EPR correlations in position-momentum variables \cite{einstein1935pr}. 
\begin{figure*}[t!]
\includegraphics[scale=1]{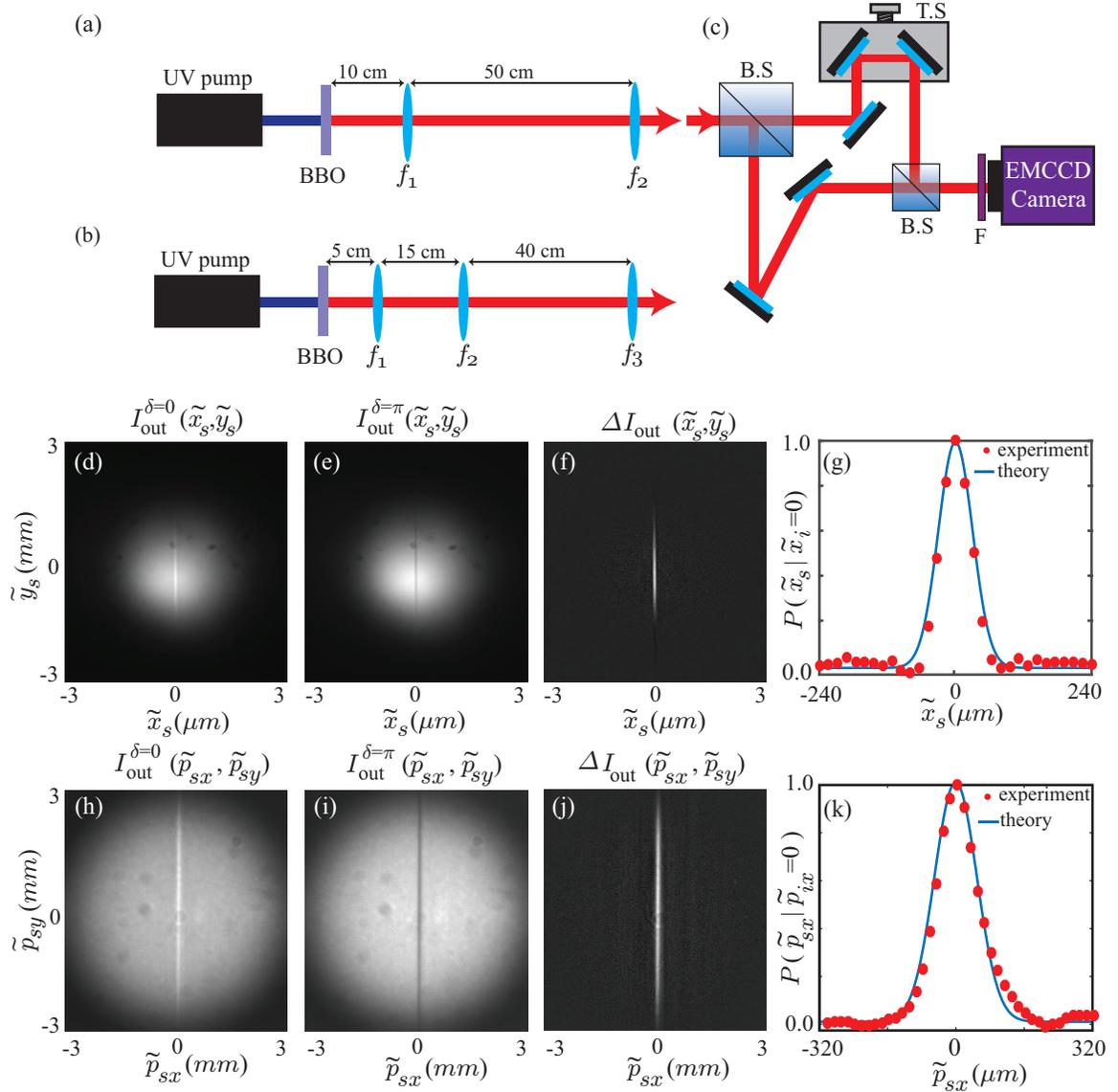}
\caption{ (a) Lens configuration for measuring position correlation. (b) Lens configuration for measuring momentum correlation. (c) Inversion-based interferometer for measuring position and momentum cross-spectral density functions. B.S: Beam Splitter, T.S: Translational Stage, F: an interference filter of $10$ nm spectral width centered at $810$ nm..  (d) and (e) The two interferograms recorded at $\delta_c=0$ and $\delta_d=\pi$ with the configuration in Fig. 1(a). (f) The difference intensity image $\Delta I(\tilde{x}_s, \tilde{y}_s)$. (g) Experimental and theoretical conditional probability distribution $P(\tilde{x}_s|\tilde{x}_{i}=0)$. (h) and (i) The two interferograms recoded at $\delta_c=0$ and $\delta_d=\pi$ with the configuration in Fig. 1(b). (j) The difference intensity image $\Delta I(\tilde{p}_{sx},\tilde{p}_{sy})$. (k) Experimental and theoretical conditional probability distribution $P(\tilde{p}_{sx}|\tilde{p}_{ix}=0)$.
}\label{fig1}
\end{figure*} 

SPDC is a nonlinear optical process in which a pump photon at higher frequency gets down-converted into two photons of lower frequencies called the signal and idler photons. In most experimental situations \cite{d2004prl, howell2004prl, zhang2019optexp, o2005prl,leach2012pra, edgar2012natcom,reichert2018scirpt, moreau2014prl, khan2006pra,maclean2018prl, leach2010science, chen2019prl, ou1992prl}, one uses a spatially coherent pump field for down-conversion. As a result, the joint state of the signal and idler photons produced by SPDC in these experimental situations remains pure and very closely resembles the state given by Eq.~(\ref{eq1}). Therefore, we experimentally demonstrate our technique with the two-photon state produced by SPDC.

For a spatially-coherent Gaussian pump with beam waist at the crystal plane, the two-photon wavefunction produced by SPDC in the momentum basis is given by \cite{dixon2012prl,edgar2012natcom,leach2012pra,schneeloch2016jop} 
\begin{align}
\psi(\bm{p}_s,\bm{p}_i)=Ae^{-\tfrac{(\bm{p}_i+\bm{p}_s)^2\sigma_p^2}{4\hbar^2}}e^{-\tfrac{|\bm{p}_i-\bm{p}_s|^2\sigma_-^2}{4\hbar^2}},\label{mom-2}
\end{align} 
where $\bm{p}_s \equiv (p_{sx},p_{sy})$ and $\bm{p}_s \equiv (p_{ix},p_{iy})$ are the transverse momenta at the crystal plane of the signal and idler photons, respectively. $\sigma_p$ is the pump beam waist and $\sigma_-=\sqrt{{0.455L\lambda_p}/{2\pi}}$, where $L$ is the length of the crystal, $\lambda_p$ is the pump wavelength and A is a normalization constant. By taking the Fourier transform of the wavefunction given in Eq.~(\ref{mom-2}), we write the two-photon wavefunction in the position basis as 
\begin{align}
\psi(\bm{\rho_s},\bm{\rho_i}) =A'e^{-\tfrac{(\bm{\rho_i}+\bm{\rho_s})^2}{4\sigma_p^2}}e^{-\tfrac{|\bm{\rho_i}-\bm{\rho_s}|^2}{4\sigma_-^2}}.\label{pos-2}
\end{align} 
Here, $A'$ is a normalization constant, $\bm{\rho_s}\equiv(x_s,y_s)$ and $\bm{\rho_i}\equiv(x_i,y_i)$ are the transverse position vectors of the signal and idler photons at the crystal plane. We note that the above wavefunctions $\psi(\bm{p}_s,\bm{p}_i)$ and $\psi(\bm{\rho_s},\bm{\rho_i})$ satisfy the conditions given in Eqs. (\ref{eq5}) and (\ref{eq7}), respectively.

Next, we present our experiment results demonstrating how the conditional position and momentum uncertainties can be obtained by measuring the cross-spectral density functions of just the signal photon. 
Figures \ref{fig1}(a)-\ref{fig1}(c) show the schematics of our experimental setup. An ultraviolet (UV) Gaussian pump beam of wavelength $\lambda_p=405$ nm and beam waist $\sigma_p=388$ $\mu$m is incident on a $2$ mm thick $\beta-$barium borate (BBO) crystal and produces two-photon state using SPDC with the type-I collinear phase-matching. Figure \ref{fig1}(c) shows an inversion-based interferometer that we use for measuring the cross-spectral density functions \cite{bhattacharjee2018apl, kulkarni2017natcom}. Figures \ref{fig1}(a) and \ref{fig1}(b) show the lens configurations for imaging, respectively, the crystal plane and the Fourier plane of the crystal onto an EMCCD camera having $512\times 512$ pixels and $60$ second acquisition time.

For measuring the position cross-spectral density function of the signal photon, we use the configuration of Fig.~\ref{fig1}(a) with $f_1=10$ cm and $f_2=40$ cm and image the crystal onto the EMCCD plane, kept at $40$ cm from $f_2$, with a magnification $M=4$. We take $(x_s, y_s)$ and $(\tilde{x}_s, \tilde{y}_s)$ to be the position coordinates at the crystal plane and at the EMCCD plane, respectively. The two sets of coordinates are related as $\tilde{x}_s = Mx_s$ and
$\tilde{y}_s = My_s$. The intensity $I^{\delta}_{\rm{out}}(\tilde{x}_s,\tilde{y}_s)$ of the output interferogram at the EMCCD plane is given by $I^{\delta}_{\rm{out}}(\tilde{x}_s,\tilde{y}_s)=k_1I(\tilde{x}_s,\tilde{y}_s) + k_2 I(-\tilde{x}_s,\tilde{y}_s)+ 2\sqrt{k_1k_2}W(\tilde{x}_s,\tilde{y}_s,-\tilde{x}_s,\tilde{y}_s)\cos\delta$ \cite{kulkarni2017natcom}. Here, $k_1$ and $k_2$ are the scaling constants, while $k_1I(\tilde{x}_s,\tilde{y}_s)$ and $k_2I(-\tilde{x}_s,\tilde{y}_s)$ are the intensities at the EMCCD plane  coming through the two arms of the interferometer. The quantity $\delta$ is the phase difference between the two interferometric arms. If we take two interferograms $I_{\rm {out}}^{\delta_c}(\tilde{x}_s,\tilde{y}_s)$ and $I_{\rm{out}}^{\delta_d}(\tilde{x}_s,\tilde{y}_s)$ at $\delta=\delta_c$ and $\delta=\delta_d$, respectively, then it can be shown that the difference intensity $\Delta I_{\rm{out}}(\tilde{x}_s,\tilde{y}_s)=I_{\rm{out}}^{\delta_c}(\tilde{x}_s,\tilde{y}_s)-I_{\rm{out}}^{\delta_d}(\tilde{x}_s,\tilde{y}_s)$ is proportional to the position cross-spectral density, that is, $\Delta I_{\rm out}(\tilde{x}_s,\tilde{y}_s) \propto W(\tilde{x}_s,\tilde{y}_s,-\tilde{x}_s,\tilde{y}_s)$ \cite{bhattacharjee2018apl,kulkarni2017natcom}. Figures~\ref{fig1}(d) and ~\ref{fig1}(e) show the two experimentally measured interferograms at $\delta=\delta_c\approx 0$ and $\delta=\delta_d\approx \pi$, respectively, and Fig.~\ref{fig1}(f) shows the difference intensity $\Delta I_{\rm out}(\tilde{x}_s,\tilde{y}_s)$.  
From Eq.~(\ref{eq7new}), we have that $W(\tilde{x}_s, \tilde{y}_s, -\tilde{x}_s, -\tilde{y}_s)$, is proportional to the conditional position probability distribution function of the signal photon, that is, $P(\tilde{x}_s,\tilde{y}_s|\tilde{x}_i=0, \tilde{y}_i=0) \propto W(\tilde{x}_s, \tilde{y}_s, -\tilde{x}_s, -\tilde{y}_s)$. Therefore, we obtain the one-dimensional conditional position probability distribution function $P(\tilde{x}_s |\tilde {x}_i=0)$ by averaging $\Delta I_{\rm out}(\tilde{x}_s,\tilde{y}_s)$ over the $\tilde{y}_s$-direction and plotting it in Fig.~\ref{fig1}(g). 
Using Eq.~(\ref{pos-2}) and the relevant experimental parameters, we calculate the theoretical $P(\tilde{x}_s|\tilde{x}_i = 0)$ and plot it in Fig.~\ref{fig1}(g) (solid curve). We scale the $P(\tilde{x}_s|\tilde{x}_i = 0)$ plots in Fig. 1(g) such that the maximum value is one. We fit the experimental $P(\tilde{x}_s|\tilde{x}_i = 0)$ with a Gaussian function and find the standard deviation to be $26.25$ $\mu$m. The standard deviation of the theoretical plot is $30.6$ $\mu$m. Now, we use $\tilde{x}_s = Mx_s$ and obtain the experimental and theoretical values of $\Delta(x_s | x_i = 0)$ to be $6.56$ $\mu$m and $7.65$ $\mu$m, respectively.

For measuring the momentum cross-spectral density function of the signal photon, we use the configuration of Fig.~\ref{fig1}(b) with $f_1=5$ cm, $f_2=10$ cm, and $f_3=30$ cm. The effective focal length of this combination is $f_{e}=15$ cm. The EMCCD is kept at $30$ cm from $f_3$, which is the Fourier plane of this configuration. We take $(p_{sx}, p_{sy})$ to be the transverse momentum at the crystal plane and $(\tilde{p}_{sx}, \tilde{p}_{sy})$ is the position coordinate at the EMCCD plane. These coordinates can be shown to be related as $p_{sx}=\frac{k_0\hbar}{f_e}\tilde{p}_{sx}$ and $p_{sy}=\frac{k_0\hbar}{f_e}\tilde{p}_{sy}$ \cite{edgar2012natcom}, where $k_0=\frac{2\pi}{\lambda_0}$. The intensity $I_{\rm{out}}^\delta(\tilde{p}_{sx},\tilde{p}_{sy})$ of the output interferogram at the EMCCD plane in this case can be written as $I_{\rm out}^{\delta}(\tilde{p}_{sx},\tilde{p}_{sy})=k_1I(\tilde{p}_{sx},\tilde{p}_{sy}) + k_2I(-\tilde{p}_{sx},\tilde{p}_{sy})+ 2\sqrt{k_1k_2}W(\tilde{p}_{sx},\tilde{p}_{sy},-\tilde{p}_{sx},\tilde{p}_{sy})\cos\delta$  \cite{kulkarni2017natcom}. Here, $k_1I(\tilde{p}_{sx},\tilde{p}_{sy})$, and $k_2I(-\tilde{p}_{sx},\tilde{p}_{sy})$ are the intensities at the EMCCD plane coming through the two interferometric arms.  Just as discussed above, the difference intensity $\Delta I_{\rm out}(\tilde{p}_{sx},\tilde{p}_{sy})=I_{\rm out}^{\delta_c}(\tilde{p}_{sx},\tilde{p}_{sy})-I_{\rm out}^{\delta_d}(\tilde{p}_{sx},\tilde{p}_{sy})$ is proportional to the cross-spectral density function $W(\tilde{p}_{sx},\tilde{p}_{sy},-\tilde{p}_{sx},\tilde{p}_{sy})$  \cite{bhattacharjee2018apl,kulkarni2017natcom}. Figures~\ref{fig1}(h) and \ref{fig1}(i) show the two experimentally measured interferograms at $\delta=\delta_c\approx 0$ and $\delta=\delta_d\approx \pi$, respectively, and Fig.~\ref{fig1}(j) shows the difference intensity $\Delta I_{\rm out}(\tilde{p}_{sx},\tilde{p}_{sy})$. From Eq.~(\ref{eq7}), we have that $P(\tilde{p}_{sx}, \tilde{p}_{sy} | \tilde{p}_{ix}=0, \tilde{p}_{iy}=0) \propto W(\tilde{p}_{sx}, \tilde {p}_{sy}, -\tilde{p}_{sx}, -\tilde{p}_{sy})$. Therefore, we obtain the one-dimensional conditional probability distribution function $P(\tilde{p}_{sx}| \tilde{p}_{ix}=0)$ by averaging $\Delta I_{\rm out}(\tilde{p}_{sx}, \tilde{p}_{sy})$, over $\tilde{p}_{sy}$-direction and plotting it in Fig.~\ref{fig1}(k). Using Eq.~(\ref{mom-2}) and the relevant experimental parameters, we calculate the theoretical $P(\tilde{p}_{sx}|\tilde{p}_{ix} = 0)$ at the EMCCD plane and plot it in Fig.~\ref{fig1}(k) (solid curve).  We scale the $P(\tilde{p}_{sx}|\tilde{p}_{ix} = 0)$  plots in Fig.~\ref{fig1}(k) such that the maximum value is one. We fit the experimental $P(\tilde{p}_{sx}|\tilde{p}_{ix} = 0)$ with a Gaussian function and find the standard deviation to be $49.5$ $\mu$m. The standard deviation of the theoretical plot is $49.8$ $\mu$m. Using $p_{sx}=\frac{k_0\hbar}{f_e}\tilde{p}_{sx}$, we obtain the experimental and theoretical values of $\Delta(p_{sx}|p_{ix}=0)$ to be $2.55\times10^{-3}\hbar$ $\mu$m$^{-1}$ and $2.57\times10^{-3}\hbar$ $\mu$m$^{-1}$, respectively.

As defined in Eq.~(\ref{Heisenberg-4}), the experimentally measured value of the conditional uncertainty product $U_{\rm ex}$ is $1.67\times10^{-2}\hbar$.  This is much smaller than $0.5 \hbar$ and thus implies a strong EPR correlations between the two entangled photons. We find the theoretical conditional uncertainty product $U_{\rm th}$ to be $1.96\times10^{-2}\hbar$, and we thus find a good match between the theory and experiments. Now, in order to quantify the accuracy of our measurement scheme, we use the quantity $\mathcal{F} \equiv \frac{|U_{\rm th}-U_{\rm ex}|}{U_{\rm th}}\times 100\%$. We note that a smaller value of $\mathcal{F}$ implies better accuracy of EPR correlation measurements. For our experimental results, we obtain $\mathcal{F}=14.7\%$. This value of $\mathcal{F}$ obtained through our measurement scheme is much smaller than the previously reported values of $27.1\%$ in Ref.~\cite{moreau2014prl}, $43.7\%$ in Ref.~\cite{edgar2012natcom}, $66\%$ in Ref.~\cite{howell2004prl}, $190\%$ in Ref.~\cite{dixon2012prl}, and $376\%$ in Ref.~\cite{leach2012pra}. 
There is another quantity that is quite often used for quantifying the EPR-correlations measurements. It is called the degree of violation and is defined as $\mathcal{D}=\left({0.5 \hbar}/{U_{exp}}\right)^2$. The degree of violation $\mathcal{D}$ does not quantify the measurement accuracy but gives an estimate of the degree with which the Heisenberg bound of $0.5\hbar$ is violated. For our experimental measurements, $\mathcal{D}=896$, whereas the degree of violations reported earlier include $576$ in Ref.~\cite{moreau2014prl}, $380$ in Ref.~\cite{edgar2012natcom}, $42$ in Ref.~\cite{dixon2012prl}, $25$ in Ref.~\cite{howell2004prl}, and $4$ in Ref.~\cite{leach2012pra}. Thus we report not only the most accurate EPR-correlations measurements but also the highest degree of violation. We note that since we are using collinear phase-matching, what gets recorded by the EMCCD camera is the sum of the interferograms produced by the signal and idler fields. However, since signal and idler photons are identical in their spatial degree of freedom, functional form of the sum interferogram is same as that of the individual interferograms produced by signal and idler photons.


In conclusion, we have demonstrated a scheme for measuring two-photon position-momentum EPR correlations that does not require coincidence detection. Our scheme works for any two-photon state that is pure, irrespective of whether the state is separable or entangled. We have experimentally demonstrated this technique with pure two-photon states produced by type-I SPDC and have obtained the most accurate measurement of position-momentum EPR correlations reported so far. Our scheme can be extended for measuring EPR-correlations in other continuous variables such as time-energy \cite{khan2006pra,maclean2018prl,mei2020prl} and angle-OAM \cite{leach2010science}. It is also applicable to pure two-particle states produced using other processes such as spontaneous four-wave mixing \cite{lee2016prl}. Thus, we expect our work to have wide practical implications for continuous-variable quantum information applications.

We thank Siddharth Ramachandran for useful discussions and acknowledge financial support through the research grant no. EMR/2015/001931 from the Science and Engineering Research Board (SERB), Department of Science and Technology, Government of India, and the research grant no. DST/ICPS/QuST/Theme-1/2019 from the Department of Science and Technology, Government of India. NM acknowledges post-doctoral fellowship from Indian Institute of Technology Kanpur.

\bibliographystyle{apsrev4-2}
\bibliography{ref}

\end{document}